\documentstyle[preprint,prb,aps]{revtex}

\begin{document}

\title{Study of $Ce$ intermetallic compounds: an LDA classification and hybridization effects}

\author{V. L. Vildosola and A. M. Llois}

\address{Departamento de F\'{\i}sica, Comisi\'on Nacional de Energ\'{\i}a
At\'omica, Avenida Libertador 8250, 1429 Buenos Aires, Argentina}

\date{\today}
\maketitle

\tightenlines

\begin{abstract}
Spin-polarized calculations within the LDA approximation have been done with the
goal of characterizing $Ce$ intermetallic compounds. Symmetry and chemical
environment effects on $4f$ hybridization and, thereafter, on magnetization effects
have been studied.
\end{abstract}

\section{Introduction}

In the last decades, $Ce$ intermetallic compounds have received a lot of attention
both in theory and experiment. On one side, these systems can exhibit a large variety
of behaviors, such as superconductivity, huge magnetoresistance, itinerant 
magnetism, etc. On the other side, the nature of the $4f$ states is controversial
since it has not been established yet which systems can be treated using a 
bandlike picture and which ones using a localized one. In the first picture, 
the effective electronic correlation is weak and the $4f$ states form a narrow 
band\cite{mackintosh}, while in the second one, the $4f$ electrons are highly correlated and
interact weakly with the conduction band\cite{anderson}.
   
The main difference between $Ce$ and other rare earth systems is that, in the case
of Ce, the $4f$ state is energetically very close to the Fermi level so that its occupation 
number and the strength of hybridization with 
the conduction electron bands strongly depend on the chemical and geometrical 
environment. In that sense , we can understand the complexity of the phase 
diagram of pure metallic $Ce$ which, depending on temperature and
pressure, can be magnetic, paramagnetic or superconducting\cite{elliott}.
Due to this fact, different types of magnetism have been observed
within the intermetallic compounds: intermediate valence behavior, Kondo effect
and magnetic ordered structures (FM, AF) are among them\cite{kappler}.
 Intermediate valence behavior appears in systems having the states with
 $n$ and $n-1$ $4f$ electrons almost degenerate.    
The nearly degenerate condition is a consequence of hybridization of the $4f$ states
 and results in a configuration with a non integer electronic occupation number 
 $n_{4f}$. In general, 
intermediate valence materials show anomalous values for some physical 
properties such as lattice parameter, bulk modulus, magnetic 
susceptibility, etc, as compared to systems with an integer $4f$ occupation
number. The intermediate valence behavior in $Ce$ compounds is characterized 
 by having a very small or zero magnetic moment\cite{lawrence}. 
 
The different physical situations posed by these systems have been
traditionally described by many body hamiltonians which consider strong 
electron-electron interactions and treat hybridization as a small 
perturbation. Kondo systems, for which correlations are dominant, have to be treated with these model 
hamiltonians. In $Ce$ intermetallics, particularly, it is not clear  at all how important the
relative strength of hybridization and correlations are. There are systems whose electronic 
ground state can be well described by an itinerant picture, namely those in
which $4f$ hybridization plays a relevant role giving rise to a decrease in
correlation effects\cite{cenix}. 
 
In this work our aim is twofold. First we want to use the band picture to establish a criterion for 
characterizing the ground
state of $Ce$ intermetallic compounds by analysing their LDA spin contribution to
the magnetic moment. As a second goal, and using for this the established
criterion as a tool, we also want to analyse the dependence of the
magnetism of $Ce$ on geometrical and chemical environment. 
For this we study a variety of $Ce$ compounds, whose magnetic
properties are experimentally well known. We claim that it is possible to
determine to which regime the system under study belongs. That is, if we are dealing 
with strongly hybridized or strongly localized systems, going
through the intermediate situations, by analysing the itinerant magnetic contribution that results from spin
polarized LDA calculations. This analysis is certainly done in
the knowledge that only those $Ce$ compounds whose $4f$ states are essentially itinerant can be well
described within the LDA band theory frame. Actually, a proper treatment for the $4f$
electrons would be to include the self-interaction correction (SIC)\cite{valency} in the calculations but this is not
necessary for the kind of description undertaken in this work.
 
Once we have established our characterization tool, we discuss the relative
importance of chemical and crystalline environment on hybridization and
thereafter on the determination of the magnetic state of intermetallic
systems of the type $CeX_{n}$ with $n$ equal or larger than one.

This paper is organized as follows. In Sec. II we report the results of
calculations done for several real systems and present a discussion of how the LDA approximation 
 can be used to classify them into magnetic, intermediate valence or
paramagnetic compounds. 
In the second part of this section we analyse the influence of chemical and structural environment on
the magnetic ground state of $Ce$ compounds. We finally present the conclusions
 in Sec. III.

\section{Electronic structure calculations}

In this work we perform {\it ab-initio} calculations using the Full Potential
Linearized Augmented Plane Waves (FP-LAPW) method in 
the Local Density Approximation (LDA)\cite{wien}. We use the exchange and correlation  
potential of J.P. Perdew and Y. Wang\cite{par}. We make both paramagnetic and
 spin polarized calculations. The
sampling of the Brillouin zone used to calculate the electronic ground state  
depends on the size and symmetry of each system. In general, from 800 to 1400 
k-points in the first Brillouin zone are enough for convergence. The  considered {\it muffin tin}
 radios, $R_{mt}$, are equal to
2.8 au for Ce, 2.4 au for $4d$ transition metals (TM), 2.0 au for $3d$ transition metals (TM), 
1.8 au for S and 1.6 au for N. The cutoff parameter which gives the number of plane waves in the
interstitial region
is taken as $R_{mt}K_{max}=8$, where $K_{max}$ is the maximun value of the reciprocal 
lattice vector used in the expansion of plane waves in that zone.
The total energy is converged up to $10^{-4} Ry$.

\subsection{ Characterization of $Ce$ compounds through their LDA magnetic moments} 

It is well known that $Ce$ has one $4f$ electron in the solid. As we have
mentioned before, the energy of this $4f$ state is very close to the 
Fermi level so that it is, in principle, very sensitive to chemical and geometrical 
environment. In some compounds $Ce$ keeps its magnetic moment equal to 1 $\mu_{B}$ and in 
others it can decrease even going to zero. In this Section we establish a
way of
characterizing $Ce$ compounds which allows us to classify them into magnetic,
intermediate valence or paramagnetic by doing LDA spin polarized calculations. 
 To achieve this we study compounds
 whose magnetic properties have already been reported and are listed in Table I.

 $CeNi$, $CeRh$ and $CePd_{3}$ are accepted to be intermediate valence compounds\cite{sereni2} 
 while $CeN$ and $CeRh_{3}$ are well described by an itinerant picture\cite{delin}, 
 being paramagnetic. $CeS$, $CePd$, $CeAg$
 and $CeCd$ are magnetically ordered systems. Table I contains the structural data
 and the experimental ordering temperatures, $T_{C}$ and $T_{N}$, which correspond 
 to the Curie and Ne\`{e}l temperatures depending on whether the compounds are 
 ferromagnetic or antiferromagnetic. No magnetic order has been experimentally observed for the 
 cases where no ordering temperature is shown\cite{sereni}.   
 
In our calculations three types of magnetic configurations are possible: paramagnetic (P),
ferromagnetic (F) and antiferromagnetic(AF). In Table I we also list the more stable configuration and
the obtained LDA magnetic moment of $Ce$ for the different compounds.
The calculations are performed at the experimental volumes shown in the same table.

The obtained magnetic moments for the compounds $CeNi$, $CeRh$ and $CePd_{3}$ are far from being
1$\mu_{B}$ but they are clearly not zero so that we cannot consider them as
being paramagnetic. In that
sense it is not surprising that these compounds are those which are widely accepted to
be intermediate
valence systems. An intermediate valence system is usually defined in the literature as one having
in the average a non integer number of $4f$ electrons. 

It is very important to keep in mind that LDA calculations cannot account for the strong 
correlation effects that may occur in some rare earth compounds. However, we
can infer from their itinerant contribution and using LDA the
correct ground state for the systems studied here.

Based on the comparison of experimental data and our LDA results
, we propose that, depending on the LDA magnetic moment,  a $Ce$ compound can be considered as

\vspace{0.25in}
\begin{center}

\hspace{0.4in} Itinerant \hspace{0.3in} if \hspace{0.1in} $\mu_{Ce} = 0$,

\vspace{0.17in}

\hspace{0.05in} Intermediate valence  \hspace{0.1in} if \hspace{0.1in} $\mu_{Ce} < 0.5\mu_{B}$ 

or 

\hspace{0.5in} Magnetic  \hspace{0.3in} if \hspace{0.1in} $\mu_{Ce} > 0.5
\mu_{B}$.
\end{center}
\vspace{0.25in}
 
 The systems that are accepted to be itinerant, such as $CeN$ and $CeRh_{3}$,
are the ones whose magnetic moment is exactly equal to zero. That is, when in a given $Ce$ system 
the magnetic moment is zero the $4f$ electrons have a strong itinerant character. 

In $Ce$ compounds the degree of
localization is closely related to the strength of $Ce$ hybridization and consequently to the magnetic
state. Let us take two extreme situations as examples, $CeRh_{3}$ and $CeAg$. 
In Figure 1 we show an electronic charge density plot corresponding to one of the 
Kohn-Sham orbitals with energy close to the
Fermi level and with $90\%$ of $4f$ character. In the $CeRh_{3}$ case, there is a mixing between $Rh$ and
$Ce$ states near $E_{F}$ while, this is not the case for $CeAg$, as can be clearly seen in Figure 1.
 
 We can also analyse the degree of hybridization by comparing $4f$ and $4d$ partial densities 
of states (Figure 2). The most striking difference between $CeRh_{3}$ and $CeAg$ 
systems is that in the last one, the $4d$ and $4f$ bands are approximately 4 eV apart being both of them very narrow,  
of the order of 1eV, while in  $CeRh_{3}$, the $4d$ band is more extended in energy
 (about 4eV), leading to an energy region around the Fermi level where hybridization is important. 
 Thus, taking into account the calculated magnetic moments and using the established criterion, we
 say that in the compound where the $4f$ band is more 
hybridized, namely $CeRh_{3}$, the $4f$ band turns more delocalized leading thus
to a non magnetic Ce. This should hold in general.
  
\subsection{ Dependence of $Ce$ magnetism on crystalline and chemical environment}

$Ce$ can completely or partially loose its magnetic moment
 depending on chemical and crystalline environment. In this section we study how $4f$ hybridization
  affects the $Ce$ magnetic moment depending on the local symmetry and the chemistry of the ligand. 
  The $4f$ band hybridizes not only with $4f$ electrons of neighboring $Ce$ sites but also with orbitals 
  of the ligand ($4d$, $3d$ or $sp$ bands). Both types of hybridization
  produce a decrease in the magnetic moment of Ce.

For the symmetry considerations, we compare systems of the type $CeX$ (X=Ni, Rh, Pd) and analyse them in the CrB and CsCl
crystal structures. These structures are being taken as prototypes of different symmetries within the same relative
composition. Actually, both structures appear in nature associated with $CeX$ compounds.
In general, when X is a latter TM the observed structure is CrB and when X belongs to the
1B, 2B or the 3A column of the periodic table the corresponding structure is the CsCl one.
 
The calculation for a given $CeX$ system in both the CrB and the CsCl structures helps 
 us to understand how local symmetry affects $4f$ hybridization and consequently the 
 magnetization of Ce. 
  The calculations within the CrB structure are performed at the
experimental equilibrium volume at room temperature and, since the CsCl structures 
are hypothetical ones for $CeNi$, $CeRh$ and $CePd$, we take the same volume per atom as in 
CrB in order to be able to compare the outcoming magnetic properties.

From Table II we see that the CsCl structure favours magnetism, even if slightly,
as compared to the CrB one. This is a consequence of CsCl having higher symmetry. 
Crystal field effects lead, in the CrB case, to a lifting of almost all of the $4f$
degeneracies while this is not the case in the CsCl structure. Within a 'quasi'-Stoner 
image CsCl favours magnetism due to a higher density of states at the Fermi level. 
This
is coherent with the fact that only some of the CrB compounds are magnetic while all of the CsCl ones
are magnetic. 

Figure 3 shows charge density plots with contributions stemming from the energy range $0.7E_{F}<E<E_{F}$, where 
$E_{F}$ is the Fermi energy. In this range the $4f$ band is the most important one. The plots
show, for the $CeRh$ system, the charge densities projected into the (010) plane for CrB 
and into the (110) one for CsCl. In the CrB structure the $4f$-$4f$ hybridization between neighboring $Ce$ sites is 
more important than the $4f$-$4d$ or
$4f$-$3d$ hybridization between $Ce$ and TM atoms. This can be infered from the fact that there is more
charge between $Ce$ atoms than between $Ce$ and TM, namely a weight of 0.02 in CrB as compared to 0.007 in CsCl.
 On the other hand,
the $4f$-$d$ hybridization is stronger in CsCl than in CrB. The amount of total
interstitial charge is practically the same in both structures: CrB and CsCl structures have 
4.4 and 4.3 interstitial  electrons per formula unit respectively. The difference among them comes from 
the spatial distribution of charge.

For the systems under study which crystallize in the CsCl structure, symmetry makes a contribution 
to the magnetic moment of $Ce$, but it is
actually not the crystalline environment the determining factor for the magnetic behavior. It is rather the
chemical nature of the ligands the one that is responsible for this magnetic result. Considering for instance, the $CeAg$ or $CeCd$ 
system from the previous section, which
crystallize in the CsCl structure, the 4d band lies very low in energy and thereafter there is nearly no
hybridization betwen $4f$-$4d$ bands, leading this to the magnetization of Ce. In the hypothetical situation
that one could force these systems to crystallize in the CrB structure they would also be magnetic. 

On the other hand, if
 we compare the magnetization of $CeNi$, $CeRh$ and $CePd$ focusing on the CrB structure, which
 is the real one for these compounds, we
 can get insight into the effect that the type of ligand has on the magnetic 
 moment of $Ce$. Using our criterion $CePd$ is a magnetically ordered system, while $CeRh$ and $CeNi$ 
 are not when considered in their natural CrB structure. In Figure 4 we show the densities of states 
 for the three compounds. The different magnetic solutions within the same crystalline structure can be
explained, by the fact that in $CeRh$ the $4d$ band is closer in energy to the Fermi level
 than in the $Pd$ compound and consequently $4f$-$4d$
 hybridization is stronger.  The more the $4f$ band hybridizes with the $4d$ band the
  smaller is the $Ce$ magnetic moment.  On the other hand, in the $CeNi$ case, there is an interplay
  between two types
  of hybridization namely $Ce$-$Ce$ and $Ce$-$Ni$, both induce a decrease of the magnetism  of Ce as
  compared to $CePd$. Actually, in $CeNi$, there is a reduction in volume and $Ce$ atoms are nearer
 to each other than in CePd. Consequently, as can be seen from the densities of states in Figure 4, 
 the $4f$ band is 0.5 eV wider in $CeNi$ than in $CePd$ due to an increase in $Ce$-$Ce$ mixing. 
 On the other side, the $3d$ band lies nearer to $E_{F}$ than the $4d$ one this giving rise to $4f$-$3d$ mixing.
  Due to this interplay $CeNi$ is the less magnetic compound among the CrB systems.
 
\section{Conclusions}

In this contribution using an itinerant picture we show that it is possible to characterize the magnetic 
ground state of $Ce$ intermetallic compounds by doing {\it ab-initio} calculations within the LDA approximation. 
This criterion allows us to classify them into magnetically ordered, intermediate valence or paramagnetic systems 
trough their calculated spin contributions to the $Ce$ magnetic moment. 
It is based on the band theory frame in which, in general, correlated $Ce$ systems
are not well described. However we find that it is an usefull tool to obtaine qualitative
information about the electronic ground state of a wide variety of $Ce$ compounds, including
those with strong electronic correlations.

We study the importance of $4f$ hybridization in determining the magnetic state of $Ce$ by analysing both symmetry and
chemical effects. 
We study first the influence of the symmetry environment on $4f$ hybridization in $CeNi$, $CeRh$ and 
$CePd$ and take CsCl and CrB as prototypes of high and low symmetry structures. We see that CsCl 
slightly favour magnetism as compared to CrB. This fact can be understood with the following argument: 
CrB's local environment has less symmetry operations than CsCl (8 vs. 48) and consequently there is a 
lifting of $4f$ degeneracies giving rise to a smaller density of states at the Fermi energy. Within a 
Stoner picture there is a stronger instability for magnetism in CsCl than in CrB. In this sense we can 
understand the fact that all the systems with the formula unit CeX are magnetic when growing in the
CsCl structure (with X belonging to the 1B, 2B and
3A column of the periodic table) but they are not when growing in the CrB, in which only $CePd$ and $CePt$ 
are magnetically ordered. 

Noteworthy is the fact that, in the cases studied, local symmetry is not a determining factor for
the magnetic behavior. It is actually the type of ligand the crucial factor to determine the magnetic state.  
Along this line, we focus our study on the CrB structure to analyse the effect of chemical environment. 
We conclude that both $4f$-$d$ and $4f$-$4f$ types of hybridization can lead to a decrease of $Ce$ magnetic moment 
in the systems $CeRh$ and $CeNi$ with respect to $CePd$. In the first case it is mainly the
mixing between the $4d$ band of $Rh$ with the $4f$ one from $Ce$ what produces a strong decrease in the
magnetic moment. In the second case, $CeNi$, both types of hybridization occur, resulting this in an even lower value
of the magnetic moment of $Ce$.

\section{Acknowledgments}

We would like to thank Dr. J. G. Sereni for having encouraging the study of these systems. We also thank Dr. M.
Weissmann for helpful and fruitful discussions. We
acknowledge Consejo Nacional de Investigaciones Cient\'{\i}ficas for supporting this work. This work
was funded by ANPCyT Project No. PICT 03-00105-02043.

\begin{table} 
\begin{center}

\begin{tabular}{c|c|ccc|c|c|c}
\hline \hline
Compound & Crystal   & \multicolumn{3} {|c|} {lattice parameter (\AA)} & $T_{C}$(K)
& ground state & $\mu_{Ce}$($\mu_{B}$) \\ 
         & structure & $a$ & $b$ & $c$ &    \\ \hline                   
$CeN$    & NaCl & 5.02 &  &  & - & P & 0 \\ 
$CeS$    & NaCl & 5.79 &  &  & $8.3^{*}$ & AF & 0.61  \\
$CeNi$   & CrB  & 3.78 & 10.37 & 4.29 & - & int.val. & 0.26   \\
$CeRh$   & CrB  & 3.85 & 10.98 & 4.15 & - & int.val. & 0.42  \\ 
$CePd$   & CrB  & 3.89 & 10.91 & 4.63 & 6.6 & F & 0.83    \\ 
$CeAg$   & CsCl & 3.76 & & & 9 & F & 0.88 \\ 
$CeCd$   & CsCl & 3.86 & & & 16.5 & F & 1.01  \\
$CeRh_{3}$ & $AuCu_{3}$ & 4.02 & & & - & P & 0  \\ 
$CePd_{3}$ & $AuCu_{3}$ & 4.12 & & & - & int.val. & 0.22  \\ \hline
\end{tabular}
\end{center}
\caption{Structural data, ordering temperatures, magnetic ground state and calculated magnetic moment of $Ce$ 
for some compounds. Those systems where there is no value for
$T_{C}$ do not present magnetic order. * indicates antiferromagnetic order. P: paramagnetic. AF :antiferromagnetic order. F: ferromagnetic order.}
\label{Table I}
\end{table}

\begin{table}
\begin{center}
\begin{tabular}{|c|c|c|c|} \hline
\multicolumn{2}{|c|}{Compound} & V($\AA^{3}$) & $\mu_{Ce}$($\mu_{B}$) \\ \hline
$CeNi$   &  CrB & 21.02  & 0.26   \\ \cline{2-4}
         & CsCl & 21.02  & 0.48   \\ \hline
$CeRh$   &  CrB & 21.93  & 0.42   \\ \cline{2-4}
         & CsCl & 21.93  & 0.55   \\ \hline
$CePd$   &  CrB & 24.56  & 0.83   \\ \cline{2-4}
         & CsCl & 24.56  & 0.86   \\ \hline
\end{tabular}
\end{center}
\caption{$Ce$ magnetic moment in $CeNi$, $CeRh$ and $CePd$ systems in both CrB and CsCl
 crystal structures at the same volume per atom in the unit cell}
\label{Table III}
\end{table}

\begin{figure}
\caption{ Charge density plots corresponding to a Kohn-Sham orbital associated with
an eigenvalue close to the Fermi energy in which the $4f$ character is the most
relevant. These densities are projected into the plane (110) for the system CeAg
(above) and into the (100) one for the system $CeRh_{3}$ (below).}
\end{figure}

\begin{figure}
\caption{ $4d$ and $4f$ partial densities of state of $CeAg$ and $CeRh_{3}$ obtained from
paramagnetic calculations. The zero level corresponds to the Fermi energy}
\end{figure}

\begin{figure}
\caption{ Chage density plots for the system $CeRh$ in both CrB and CsCl
crystalline structures. Charge density contributions coming from the energy
range $0.7E_{F}<E<E_{F}$ , in which the $4f$ band is more important, are plotted. Above: 
charge densities projected into the (010) plane of the CrB
crystal structure. Below: Charge densities projected into the (110) plane of the CsCl
crystal structure.}
\end{figure}

\begin{figure}
\caption{ $4d$ and $4f$ partial densities of states of $CeNi$, $CeRh$ and $CePd$ obtained from
paramagnetic calculations. The zero level corresponds to the Fermi energy}
\end{figure}


\begin{references}

\bibitem{mackintosh}
A. R. Mackintosh, Physica B {\bf 130}, 112 (1985);W. E. Pickett and B. M. Klein,
J. Less Common Met. {\bf 93} 219 (1983)
\bibitem{anderson}
P. W. Anderson, Phys. Rev{ \bf 124}, 41 (1961); O. Gunnarsson and K. Schonhammer,
Phys. Rev. Lett. {\bf 50}(8), 604 (1983)
\bibitem{elliott}
{\it Magnetic Properties of Rare Earth metals}, edited by R. J. Elliott,
University of Oxford, Department of Theoretical Physics, Parks Road, Oxford,
Plenum Press, London and New York, 1972
\bibitem{kappler}
J. P. Kappler {\it et al.}, Physica B {\bf 171}, 346(1991); G.L. Nieva et al., Z. Phys.
B-Conde. Matt. {\bf 70}, 181(1988); S. M. M. Evans, A. K. Bhattacharjee and B.
Coqblin, Physica B {\bf 171} 293(1991).
\bibitem{lawrence}
J. Lawrence and D. Murphy, Phys. Rev. Lett. {\bf 40}, 961 (1978)
\bibitem{cenix}    
L. Nordstrom, M. S. S. Brooks and B. Johansson, Phys. Rev. B {\bf 46}, 3458 (1992); L. Severin and B. Johansson, Phys. Rev.
B {\bf 50}, 17886 (1994); O. Eriksson et al., Phys. Rev. Lett {\bf
60}, 2523 (1988)
\bibitem{valency}
A. Svane, Phys. Rev. Lett {\bf 72}, 1248 (1994); P. Strange {\it et al.}, Nature {\bf 399}, 756 (1999)
\bibitem{wien}
P. Blaha, K. Schwarz and J. Luitz, WIEN97, Vienna University of Technology
1997. (Improved and apdated Unix version of the original copyrighted WIEN-code,
which was published by P. Blaha, K. Schwarz, P. Sorantin and S. B. TTrckey in
Comp. Phys. Commun. {\bf 59}, 399 (1990).
\bibitem{par}
J. P. Perdew and Y. Wang, Phys. Rev. B {\bf 45}, 13244 (1992)
\bibitem{sereni2}
J. G. Sereni, E. Beaurepaire and J. P. Kappler, Phys. Rev. B {\bf 48}, 3747(1993); J. P. Kappler {\it et al.}, J. Less-Common
Met., {\bf 111}, 261 (1985)
\bibitem{delin}
  A. Delin {\it et al.}, Phys. Rev. B {\bf 55}, 10175 (1997); E. Weschke {\it et al.}, Phys. Rev. Lett. {\bf 69}, 1792 (1992). 
\bibitem{sereni}
J. G. Sereni, {\it Handbook on the Physics and Chemistry of Rare Earths}, vol.
15, edited by K. A. Gschneidner, Jr. and L. Eyring.
\end{references}
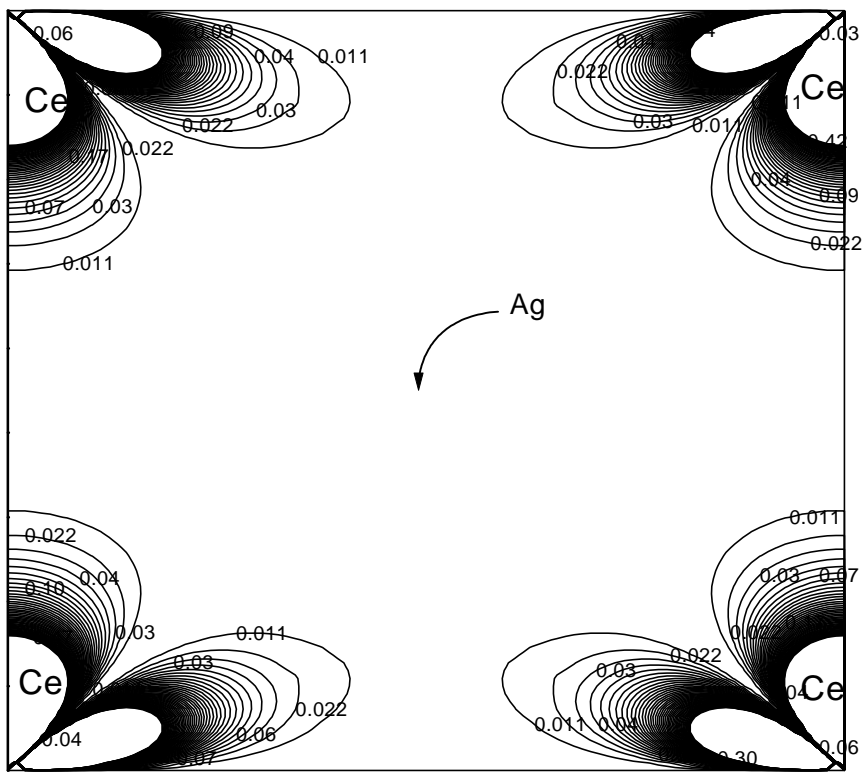
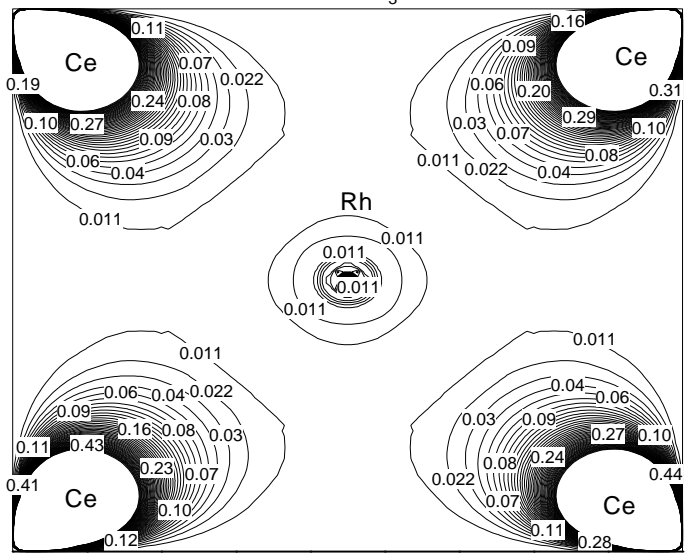
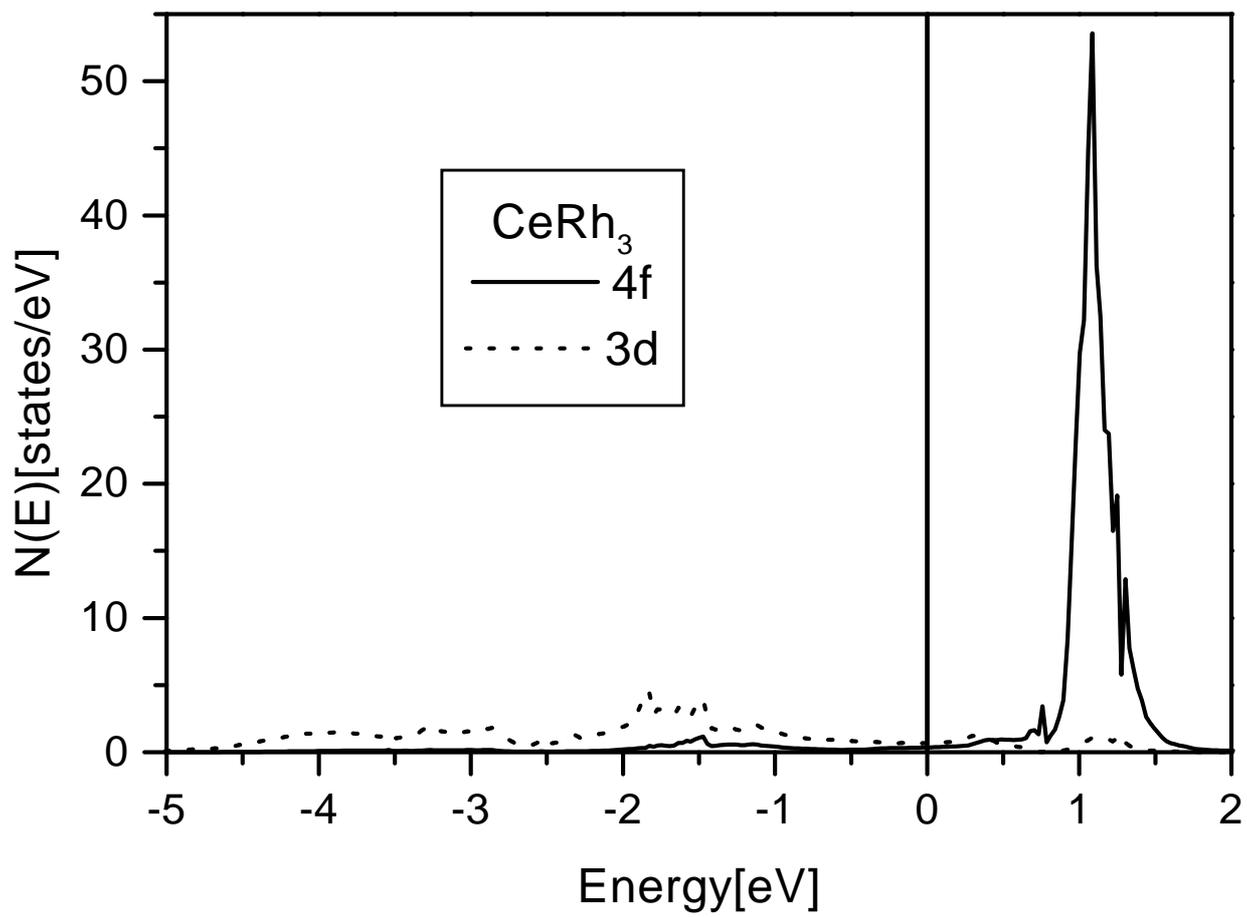
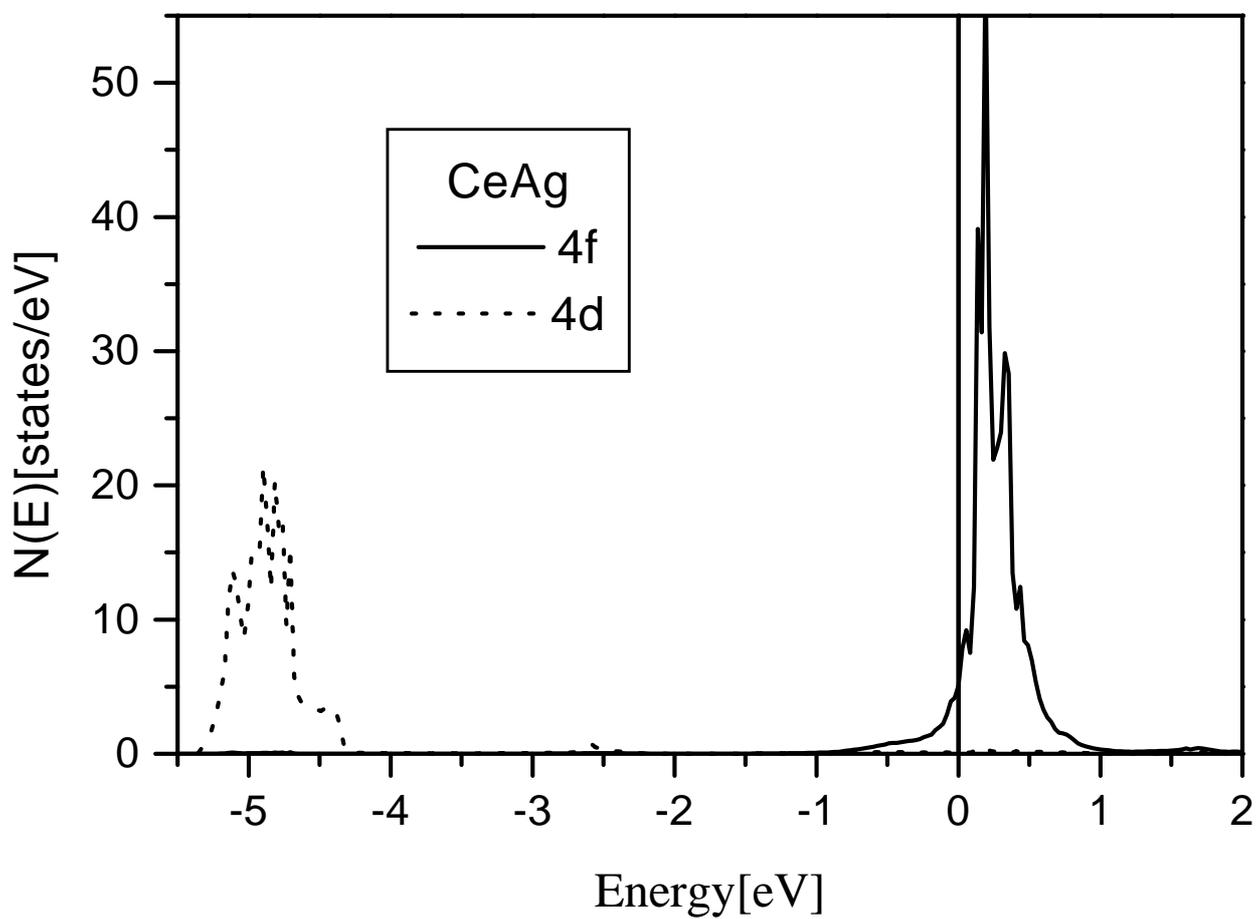
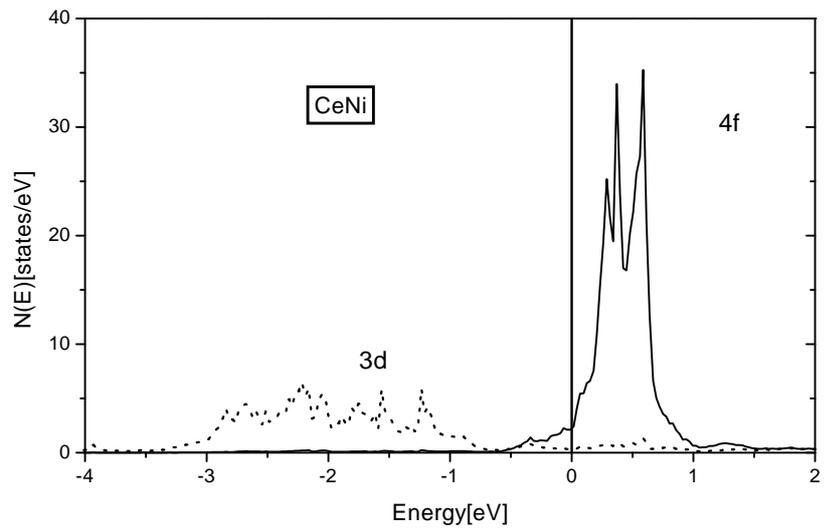
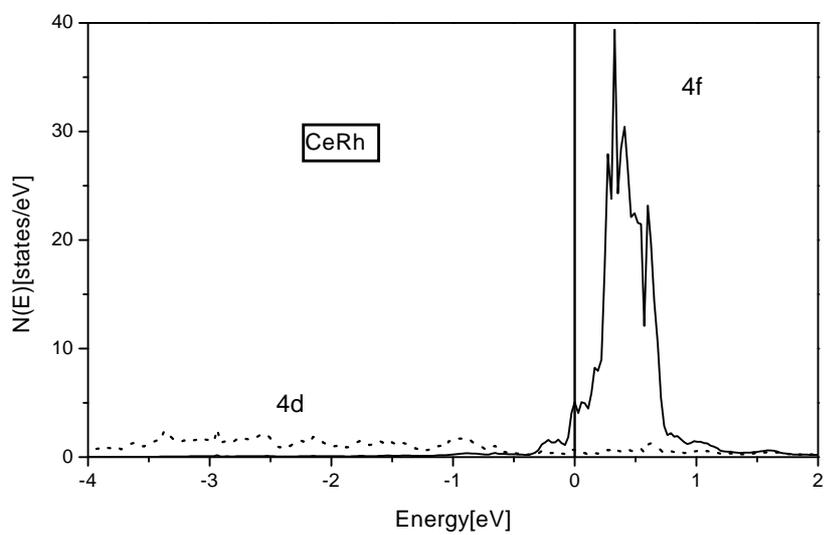
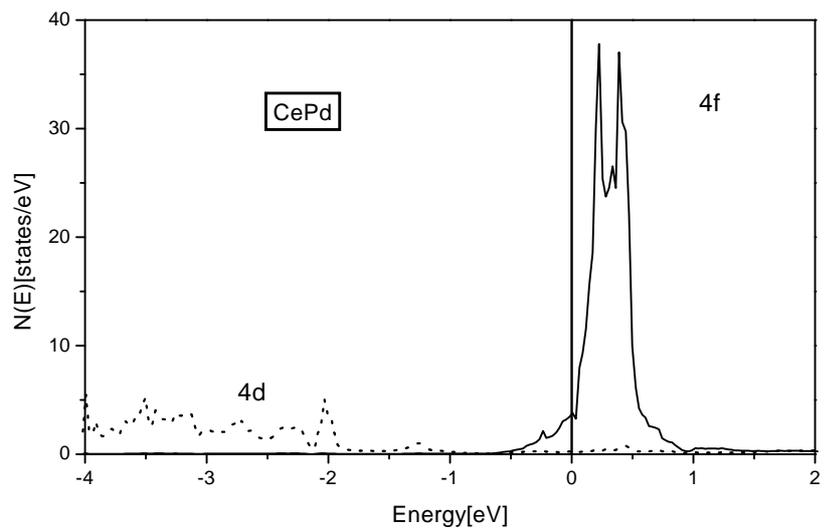
\end{document}